\documentclass[aps,twocolumn,prl,superscriptaddress,showpacs,tightenlines]{revtex4}
\usepackage{amsmath}
\usepackage{epsfig}
\usepackage{txfonts}

\begin{document}

\title{Spontaneous symmetry breaking in thermalization and
anti-thermalization}
\author{Jie-Qiao Liao}
\affiliation{Institute of Theoretical Physics, Chinese Academy of
Sciences, Beijing 100190, China}
\author{H. Dong}
\affiliation{Institute of Theoretical Physics, Chinese Academy of
Sciences, Beijing 100190, China}
\author{X. G. Wang}
\affiliation{Zhejiang Institute of Modern Physics, Department of Physics, Zhejiang
University, Hangzhou 310027, China}
\author{X. F. Liu}
\affiliation{Department of Mathematics, Peking University, Beijing
100871, China}
\author{C. P. Sun}
\affiliation{Institute of Theoretical Physics, Chinese Academy of
Sciences, Beijing 100190, China}

\begin{abstract}
The phenomenon of spontaneous symmetry breaking is investigated in
the dynamic thermalization of a degenerate quantum system. A
three-level system interacting with a heat bath is carefully studied
to this end. It is shown that the three-level system with degenerate
ground states might have different behaviors depending on the
details of the interaction with the heat bath when the temperature
approaches zero. If we introduce an external field to break the
degeneracy of the ground states and let it approach zero after
letting the temperature approach zero, then two possibilities will
arise: the steady state is a definite one of the degenerate states
independent of the initial state, or the steady state is dependent
on the initial state in a complicated way. The first possibility
corresponds to a spontaneous symmetry breaking of the system and the
second one implies that the heat bath could not totally erase the
initial information in certain cases.

\end{abstract}

 \maketitle

\narrowtext \emph{Introduction.---}Conventionally, thermalization is
understood as a dynamic process in which an open quantum system (OQS)
approaches an equilibrium state with the same temperature $T$ as that of its
heat bath~\cite{Breuer}. According to the third law of thermodynamics (the
generic version), the entropy of a non-degenerate system will vanish at the
absolute zero temperature. That is to say, at the absolute zero temperature,
the system will reach a steady pure state. Is it still the case when the
system has degenerate ground states? This turns out to be a subtle problem
related to the process of thermalization. In quantum physics, thermalization
seems to be a much more complex concept than in classical physics. In fact,
most recently a new kind of thermalization, called canonical thermalization~%
\cite{Popescu,Goldstein,Gemmer,Dong}, has been proposed.

At zero temperature a system with degenerate ground states might have a
finite entropy depending on the degree of degeneracy $d$: $S=k_{B}\ln d$~%
\cite{kersonhuang}. Roughly speaking, this can be understood by studying the
following two non-commutative limit processes: letting the perturbation
introduced to break the degeneracy approach zero and letting the temperature
of the heat bath approach zero. Actually, in the thermal equilibrium case
(the system is thermalized to reach a thermal equilibrium state), if we
first take the second limit and then take the first limit, the OQS will
reach a definite pure state and thus have a vanishing entropy. This is a
phenomenon of spontaneous symmetry breaking (SSB)~\cite{kersonhuang,ssb}. On
the other hand, if we reverse the order of these two limit processes, the
OQS will reach a maximally mixing state of all the degenerate ground states
and thus possess a non-vanishing entropy.

\begin{figure}[tbp]
\includegraphics[bb=81 350  530 674, width=6 cm]{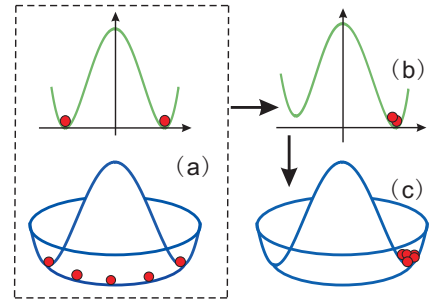}
\caption{(Color online). The schematic illustration of spontaneous symmetry
breaking (SSB) in thermalization. (a) The heat bath usually does not single
out a particular state from the degenerate ground states at zero
temperature; (b) Breaking the symmetry of the potential will cause a
preference of the system to a particular ground state at zero temperature;
(c) After the symmetry is recovered at zero temperature, the system will
remain in the preferred state.}
\label{ssb}
\end{figure}

Unfortunately, the above discussion about the thermalization of a system
with degenerate ground states proves to be overly simplified. In the study
of the dynamic thermalization of a simple system, we find that the above
mentioned SSB in the thermalization can only happen when bath induced
transition is not forbidden by some selection rule. If there is a selection
rule to forbid the bath induced transition between the degenerate ground
states, the steady thermalized state will depend on the initial state and
thus the SSB will not appear. In this case, the OQS enjoys the so called
anti-thermalization effect: some information of the initial state is kept
after the OQS is thermalized to a steady state.

We will study a three-level system interacting with a heat bath. The two
lower (or higher) energy states $|g_{1}\rangle$ and $|g_{2}\rangle$ of this
three-level system are degenerate and can be split by applying an external
field. The heat bath is modeled as the bath of harmonic oscillators. The
dynamic process of the system's approaching the steady state at zero
temperature will be carefully analyzed from the master equation approach. We
will prove that if there exists a non-vanishing coupling to the bath for
arbitrary two energy levels of the three-level system, then the third law of
thermodynamics is valid thanks to the SSB. On the other hand, for the
conventional $\Lambda$- and V-type atoms, we will reveal the exotic
anti-thermolization effect. This effect happens as a result of the absence
of the bath coupling induced quantum transition between $|g_{1}\rangle$ and $%
|g_{2}\rangle$ and the occurrence of the quantum interference between the
transition to $|g_{1}\rangle$ and the transition from $|g_{2}\rangle$. At
zero temperature, these conclusions coincide with those reached in the
context of the spontaneous emission of $V$-type atom in vacuum~\cite{Agarwal}%
.

\narrowtext \emph{SSB in thermalization.---}In general, the Hamiltonian $%
\hat{H}_{S}$ of an OQS to be thermalized can be written as $\hat{H}%
_{S}=\sum_{n,\alpha }E_{n}|n,\alpha \rangle \langle n,\alpha |$, where $%
|n,\alpha \rangle $ $(\alpha =1,2,\cdots,d_{n})$ are degenerate states
correspond to the same eigenvalue $E_{n}$ $(n=1,2,3,\cdots)$ and $d_n$ is
the degree of degeneracy. Let an external field be applied to break the
energy level degeneracy as $E_{n}\rightarrow E_{n}+\Delta _{\alpha }$ and
then let this degeneracy split system contact with a heat bath of
temperature $T$ for a time longer than the conventional relaxation time.
Then the system is supposed to be thermalized to the thermal equilibrium
state $\hat{\rho}_{S}\left(\Delta _{\alpha },\beta \right)=\sum_{n,\alpha }%
\exp[-\beta \left( E_{n}+\Delta _{\alpha }\right)]Z^{-1}|n,\alpha \rangle
\langle n,\alpha |$, where $\beta =1/(k_{B}T)$ is the inverse temperature
(hereafter, we set $k_{B}=1,\hbar =1$) and $Z=\mathtt{Tr}[\exp (-\beta \hat{H%
}_{S})]$ is the partition function of the OQS.

We observe that for $\hat{\rho}_{S}\left( \Delta _{\alpha },\beta \right) $,
there exist the following two limit processes:
\begin{align}
\lim_{\Delta _{\alpha }\rightarrow 0}\lim_{\beta \rightarrow +\infty }\hat{%
\rho}_{S}\left( \Delta _{\alpha },\beta \right) & =|1,1\rangle \langle 1,1|,
\\
\lim_{\beta \rightarrow +\infty }\lim_{\Delta _{\alpha }\rightarrow 0}\hat{%
\rho}_{S}\left( \Delta _{\alpha },\beta \right) & =\frac{1}{d_{1}}%
\sum_{\alpha =1}^{d_{1}}|1,\alpha \rangle \langle 1,\alpha |,
\end{align}
where $|1,1\rangle $ denotes the ground state with vanishing energy in the
presence of the external field. Note that taking the two limits in different
orders leads to completely different results, the former being a reflection
of the SSB phenomenon.

We would like to remark that though both of the two results are correct in
the mathematical sense, the former is physically more acceptable than the
latter, which is in accordance with the generic version of the third law of
thermodynamics. Indeed, in view of the existence of the perturbation
breaking the degeneracy, it seems reasonable to let the temperature approach
zero first in the calculation. However, things are not so simple. In fact,
due to quantum interference effect, even if we let the temperature approach
zero first in the calculation, the happening of SSB is not unconditional.
This is the main conclusion of this letter.
\begin{figure}[tbp]
\includegraphics[bb=110 347 435 569, width=6 cm]{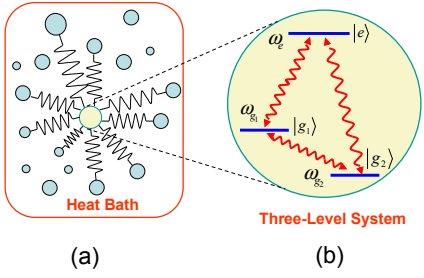}
\caption{(Color online). (a) The schematic diagram of a three-level system
immersed in heat bath, which consists of a set of harmonic oscillators. (b)
The energy level diagram of the three-level system, arbitrary two energy
levels of which couples with the heat bath.}
\label{schematicdiagram}
\end{figure}

\narrowtext \emph{Dynamic thermalization of a three-level system.---}%
Generally, the process of dynamic thermalization begins from a factorized
initial state $\hat{\rho}(0)=\hat{\rho}_{S}(0)\otimes \hat{\rho}_{B}(\beta )$%
, where $\hat{\rho}_{B}(\beta )$ is the thermal state of the heat bath. Let
the heat bath be modeled as the harmonic oscillator system with the
Hamiltonian $\hat{H}_{B}=\sum_{j}\omega _{j}\hat{a}_{j}^{\dag }\hat{a}_{j}$.
Then we have $\hat{\rho}_{B}(\beta )=\exp(-\beta \hat{H}_{B})/Z_{B}$ where $%
Z_{B}=\mathtt{Tr}[\exp (-\beta \hat{H}_{B})]$ is the partition function. The
time evolution of the total system driven by the coupling $\hat{H}_{I}$
between the system and the heat bath is determined by $\hat{U}(t)=\exp [-i(%
\hat{H} _{S}+\hat{H}_{B}+\hat{H}_{I})]$. The steady state of the system, as $%
t\rightarrow \infty$ at $T=0$, can then be obtained by calculating the
reduced density matrix $\hat{\rho}_{S}(t,\beta )=\mathtt{Tr}_{B}[\hat{U}(t)
\hat{\rho }_{S}(0)\otimes \hat{\rho}_{B}\hat{U}^{\dag }(t)]$ of the OQS ($%
\mathtt{Tr} _{B}$ stands for tracing over the heat bath).

To be specific, let us study the dynamic thermalization of a simple
three-level system. The Hamiltonian $\hat{H}_{S}=\omega _{2}\hat{\sigma}%
_{ee}+\Delta \hat{\sigma}_{g_{1}g_{1}}$ of the three-level system (as
illustrated in Fig.~\ref{schematicdiagram}) is written in terms of the flip
operators $\hat{\sigma}_{\alpha \beta }=|\alpha \rangle \langle \beta |$ ($%
\alpha ,\beta =e,g_{1},g_{2}$), where $\omega_{l}=\omega_{e}-\omega_{g_{l}}$
($l=1,2$) and $\Delta =\omega _{2}-\omega _{1}$. Here, we choose the
eigen-energy of the state $|g_{2}\rangle$ as the energy zero point. The
interaction Hamiltonian of the three-level system with its heat bath reads
\begin{eqnarray}
\hat{H}_{I}=\sum_{l=1,2}\hat{\sigma}_{eg_{l}}\hat{B}_{l}+\hat{\sigma}
_{g_{1}g_{2}}\hat{B}_{3}+h.c.,
\end{eqnarray}
where $\hat{B}_{l}=\sum_{j}\eta _{l}(\omega _{j})\hat{a}_{j}$ $(l=1,2,3)$.
For simplicity, we assume $\eta _{l}(\omega_{j})$ to be real below.

Under the Born-Markov approximation, the evolution of the reduced density
matrix of the three-level system is governed by the master equation,
\begin{equation}
\dot{\hat{\rho}}_{S}=-i[\hat{\rho}_{S},\Delta \hat{\sigma}
_{g_{2}g_{2}}]+\sum_{l=1}^{3}\mathcal{L}_{l}[\hat{\rho}_{S}]+\mathcal{L}%
_{X}[ \hat{\rho}_{S}],  \label{mastereqaution}
\end{equation}
where
\begin{widetext}
\begin{eqnarray}
\mathcal{L}_{l=1,2}[\hat{\rho}_{S}]&=&\frac{\gamma_{l}}{2}\left(\bar{n}(\omega_{l})+1\right)
\left(2\hat{\sigma}_{g_{l}e}\hat{\rho}_{S}\hat{\sigma}_{eg_{l}}-\hat{\sigma}_{ee}\hat{\rho}_{S}-\hat{\rho}_{S}\hat{\sigma}_{ee}\right)
+\frac{\gamma_{l}}{2}\bar{n}(\omega_{l})
\left(2\hat{\sigma}_{eg_{l}}\hat{\rho}_{S}\hat{\sigma}_{g_{l}e}-\hat{\sigma}_{g_{l}g_{l}}\hat{\rho}_{S}-\hat{\rho}_{S}\hat{\sigma}_{g_{l}g_{l}}\right),\nonumber \\
\mathcal{L}_{3}[\hat{\rho}_{S}]&=&\frac{\gamma_{3}}{2}\left(\bar{n}(\Delta)+1\right)
\left(2\hat{\sigma}_{g_{2}g_{1}}\hat{\rho}_{S}\hat{\sigma}_{g_{1}g_{2}}-\hat{\sigma}_{g_{1}g_{1}}\hat{\rho}_{S}-\hat{\rho}_{S}\hat{\sigma}_{g_{1}g_{1}}\right)
+\frac{\gamma_{3}}{2}\bar{n}(\Delta)
\left(2\hat{\sigma}_{g_{1}g_{2}}\hat{\rho}_{S}\hat{\sigma}_{g_{2}g_{1}}-\hat{\sigma}_{g_{2}g_{2}}\hat{\rho}_{S}-\hat{\rho}_{S}\hat{\sigma}_{g_{2}g_{2}}\right),\nonumber \\
\mathcal{L}_{X}[\hat{\rho}_{S}]&=&\left[\frac{\gamma_{12}}{2}\left(\bar{n}(\omega_{1})+1\right)+\frac{\gamma_{21}}{2}\left(\bar{n}(\omega_{2})+1\right)\right]
\left(\hat{\sigma}_{g_{1}e}\hat{\rho}_{S}\hat{\sigma}_{eg_{2}}+\hat{\sigma}_{g_{2}e}\hat{\rho}_{S}\hat{\sigma}_{eg_{1}}\right)+\frac{\gamma_{12}}{2}\bar{n}(\omega_{1})
\left(\hat{\sigma}_{eg_{1}}\hat{\rho}_{S}\hat{\sigma}_{g_{2}e}+\hat{\sigma}_{eg_{2}}\hat{\rho}_{S}\hat{\sigma}_{g_{1}e}\right.\nonumber \\
&&
\left.-\hat{\sigma}_{g_{2}g_{1}}\hat{\rho}_{S}-\hat{\rho}_{S}\hat{\sigma}_{g_{1}g_{2}}\right)+\frac{\gamma_{21}}{2}\bar{n}(\omega_{2})
\left(\hat{\sigma}_{eg_{1}}\hat{\rho}_{S}\hat{\sigma}_{g_{2}e}+\hat{\sigma}_{eg_{2}}\hat{\rho}_{S}\hat{\sigma}_{g_{1}e}
-\hat{\sigma}_{g_{1}g_{2}}\hat{\rho}_{S}-\hat{\rho}_{S}\hat{\sigma}_{g_{2}g_{1}}\right).
\end{eqnarray}
\end{widetext}
Here, the decay rates $\gamma _{l}=2\pi \varrho (\omega _{l})|\eta
_{l}(\omega _{l})|^{2}$ and $\gamma _{lm}=2\pi \varrho (\omega _{l})\eta
_{l}(\omega _{l})\eta _{m}(\omega _{l}),$\ for $\omega _{3}=\Delta $ and $%
l\neq m,m=1,2,$ depend on the mode density $\varrho (\omega )$ of the heat
bath; $\bar{n}(\omega )=1/[\exp (\beta \omega )-1]$ is the thermal average
excitation number for the boson mode of frequency $\omega $ at temperature $T
$. Note that in the master equation~(\ref{mastereqaution}), we have
neglected the Lamb shifts.

The evolution of the density matrix elements governed by the master
equation~(\ref{mastereqaution}) can be described with the optical Bloch
equation $\dot{\mathbf{X}}=\mathbf{MX}$, where  the state vector $\mathbf{%
X=X(t)}$ and the coefficient matrix $\mathbf{M}$ are respectively defined as
$\mathbf{X}=\mathbf{X}_{R}\oplus \mathbf{X}_{S}$ and $\mathbf{M}=\mathbf{R}%
\oplus \mathbf{S}$, with $\mathbf{X}_{S}=(\langle \hat{\sigma}%
_{eg_{1}}\rangle ,\langle \hat{\sigma}_{eg_{2}}\rangle )^{T}$ and
\begin{widetext}
\begin{eqnarray}\label{mmatrix}
\textbf{X}_{R}&=&((\bar{n}(\omega _{1})+1)\langle \hat{\sigma}
_{ee}\rangle -\bar{n}(\omega _{1})\langle
\hat{\sigma}_{g_{1}g_{1}}\rangle ,(\bar{n}(\omega _{2})+1)\langle
\hat{\sigma}_{ee}\rangle -\bar{n}(\omega _{2})\langle
\hat{\sigma}_{g_{2}g_{2}}\rangle ,(\bar{n}(\Delta )+1)\langle
\hat{\sigma}_{g_{1}g_{1}}\rangle -\bar{n}(\Delta )\langle
\hat{\sigma} _{g_{2}g_{2}}\rangle ,\mathtt{Re}[C], \mathtt{Im}[C])^{T},\nonumber\\
\textbf{R}&=&\left(
  \begin{array}{ccccc}
    -\gamma_{1}(2\bar{n}(\omega_{1})+1) & -\gamma_{2}(\bar{n}(\omega_{1})+1) & \gamma_{3}\bar{n}(\omega_{1}) &
    \gamma_{12}\bar{n}(\omega_{1})(\bar{n}(\omega_{1})+1)+\gamma_{21}\bar{n}(\omega_{2})(2\bar{n}(\omega_{1})+1)
    & 0 \\
    -\gamma_{1}(\bar{n}(\omega_{2})+1) & -\gamma_{2}(2\bar{n}(\omega_{2})+1) & -\gamma_{3}\bar{n}(\omega_{2}) &
    \gamma_{12}\bar{n}(\omega_{1})(2\bar{n}(\omega_{2})+1)+\gamma_{21}\bar{n}(\omega_{2})(\bar{n}(\omega_{2})+1) & 0 \\
    \gamma_{1}(\bar{n}(\Delta)+1) & -\gamma_{2}\bar{n}(\Delta) & -\gamma_{3}(2\bar{n}(\Delta)+1) &
    \gamma_{12}\bar{n}(\Delta)\bar{n}(\omega_{1})-\gamma_{21}(\bar{n}(\Delta)+1)\bar{n}(\omega_{2}) & 0 \\
    \frac{\gamma_{12}}{2} & \frac{\gamma_{21}}{2} & 0 & R_{44} & \Delta \\
    0 & 0 & 0 & -\Delta &R_{55}  \\
  \end{array}
\right),\nonumber\\
\textbf{S}&=&\left(
               \begin{array}{cc}
                 -\frac{1}{2}[\gamma_{1}(2\bar{n}(\omega_{1})+1)+\gamma_{2}(\bar{n}(\omega_{2})+1)
    +\gamma_{3}(\bar{n}(\Delta)+1)] & -\frac{\gamma_{21}}{2}\bar{n}(\omega_{2}) \\
                i\Delta-\frac{1}{2}[\gamma_{2}(2\bar{n}(\omega_{2})+1)+\gamma_{1}(\bar{n}(\omega_{1})+1)
    +\gamma_{3}\bar{n}(\Delta)] & -\frac{\gamma_{12}}{2}\bar{n}(\omega_{1}) \\
               \end{array}
             \right),
\end{eqnarray}
\end{widetext}where $C=\langle \hat{\sigma}_{g_{2}g_{1}}\rangle $ and $%
R_{44}=R_{55}=-[\gamma _{1}\bar{n}(\omega _{1})+\gamma _{2}\bar{n}(\omega
_{2})+\gamma _{3}(2\bar{n}(\Delta )+1)]/2$. We can prove $\mathbf{M}$ to be
negative-definite or have vanishing determinant. Thus the optical Bloch
equation can possess steady state solutions.

We first consider the thermalization of the three-level system at finite
temperature $T\neq 0$. In this case, $\det (\mathbf{M})\neq 0$, thus the
steady state solution of the optical Bloch equation is $\mathbf{X}=0$, or
\begin{eqnarray}
\frac{\langle \hat{\sigma}_{ee}\rangle _{ss}}{\langle \hat{\sigma}
_{g_{l}g_{l}}\rangle _{ss}}& =e^{-\beta \omega _{l}},\hspace{0.5cm}\frac{
\langle \hat{\sigma}_{g_{1}g_{1}}\rangle _{ss}}{\langle \hat{\sigma}
_{g_{2}g_{2}}\rangle _{ss}}=e^{-\beta \Delta },  \notag \\
\langle \hat{\sigma}_{g_{2}g_{1}}\rangle _{ss}& =\langle \hat{\sigma}
_{eg_{1}}\rangle _{ss}=\langle \hat{\sigma}_{eg_{2}}\rangle _{ss}=0
\label{Boltzmannrelation}
\end{eqnarray}
for $l=1,2$. Here $\langle \hat{A}\rangle _{ss}=\mathtt{Tr}_{S}[\hat{A}\hat{%
\rho}_{S}(\infty )]$, $\hat{A}$ being the operator concerned. Considering
the normalization condition $\langle \hat{\sigma}_{g_{1}g_{1}}\rangle
_{ss}+\langle \hat{\sigma }_{g_{2}g_{2}}\rangle _{ss}+\langle \hat{\sigma}%
_{ee}\rangle _{ss}=1$, from Eq.~(\ref{Boltzmannrelation}) we then have
\begin{equation}
\langle \hat{\sigma}_{g_{l}g_{l}}\rangle _{ss}=\frac{e^{\beta \omega _{l}}}{
1+e^{\beta \omega _{1}}+e^{\beta \omega _{2}}},
\end{equation}
for $l=1,2$.

Next, we consider the case with vanishing bath temperature. In this case, we
have $\bar{n}(\omega _{1})=\bar{n}(\omega _{2})=\bar{n}(\Delta )=0$ and thus
$\det (\mathbf{M})=0$. We thus cannot obtain the steady state solution of
the optical Bloch equation by simply setting $\dot{\mathbf{X}}=\mathbf{0}$
as the resulted equation $\mathbf{0}=\mathbf{MX}$ would not have a unique
solution. Instead, we have to turn to obtain the transient solution first
and then consider the long time behavior. One can expect that it cannot be
determined without regard to the details of the interaction or the initial
state. Indeed, complexity will arise here. Let us focus on the case with $%
\gamma _{3}\neq 0$ in this section, and leave the case with $\gamma _{3}= 0$
to the next section.

When $\gamma _{3}\neq 0$, namely, there exists a bath induced coupling
between the states $|g_{1}\rangle $ and $|g_{2}\rangle $, the transient
solution, which is not presented here for the technicalities, results in $%
\langle \hat{\sigma}_{ee}\rangle _{ss}=\langle \hat{\sigma}
_{g_{1}g_{1}}\rangle _{ss}=0$, and $\langle \hat{\sigma}_{g_{2}g_{2}}\rangle
_{ss}=1$. This implies that whether the three-level system has degenerate
ground states or not, the steady state of the master equation~(\ref%
{mastereqaution}) is just the thermal equilibrium state ($\propto \exp
(-\beta \hat{H}_{S})$) of the OQS even at zero temperature. This conforms to
the conventional idea. However, when $\gamma _{3}= 0$, in the next section
we will see an exotic nature of thermalization as $T\rightarrow 0$.

Now it is easily seen that taking the limits in different orders leads to
the following different results: $\lim_{\Delta\rightarrow 0}\lim_{\beta
\rightarrow +\infty }\langle \hat{\sigma}_{g_{1}g_{1}}\rangle
_{ss}=0,\lim_{\Delta \rightarrow 0}\lim_{\beta \rightarrow +\infty }\langle
\hat{\sigma}_{g_{2}g_{2}}\rangle _{ss}=1$, and $\lim_{\beta \rightarrow
+\infty }\lim_{\Delta \rightarrow 0}\langle \hat{\sigma}_{g_{1}g_{1}}\rangle
_{ss}=\lim_{\beta \rightarrow +\infty }\lim_{\Delta \rightarrow 0}\langle
\hat{\sigma}_{g_{2}g_{2}}\rangle _{ss}=1/2 $. Following the first procedure,
we will reach the conclusion that the final steady state is the pure state $%
|g_{2}\rangle $ and that there exists SSB effect in the thermalization . But
if we adopt the second procedure, we should conclude that the final steady
state is the maximally mixing state $(|g_{1}\rangle\langle
g_{1}|+|g_{2}\rangle \langle g_{2}|)/2$. By the way we remark that the SSB
can also be seen from the von Neumann entropy $S\equiv -\mathtt{Tr}_{S}\left[%
\hat{\rho}_{S}(\infty) \ln \hat{\rho}_{S}(\infty)\right]$ of the steady
state $\hat{\rho}_{S}(\infty)$ of the three-level system. In Fig.~\ref%
{entropy}, we plot the von Neumann entropy $S$ as a function of $\Delta $
and $T$. In the figure the character of double values of $S$ at the point $%
(\Delta ,T)=(0,0)$ is clearly illustrated: along the route ($T=0$, $\Delta
\rightarrow 0$), the von Neumann entropy $S\rightarrow 0$ while along the
route ($\Delta =0,T\rightarrow 0$) the von Neumann entropy $S\rightarrow 1$.
\begin{figure}[tbp]
\includegraphics[width=7 cm]{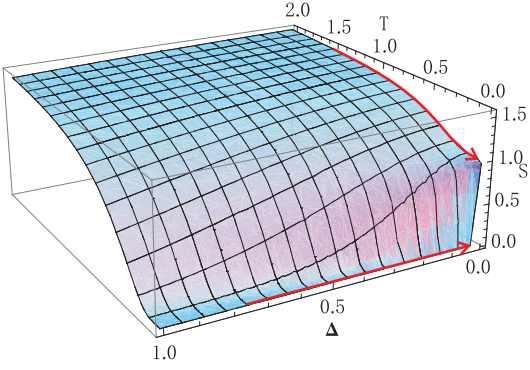}
\caption{(Color online).The von Neumann entropy of the steady state density
matrix for the three-level system is plotted versus the temperature $T$ and
the energy difference $\Delta $ between the states $|g_{1}\rangle $ and $%
|g_{2}\rangle $. The two red arrows indicate its multi-value feature
as both $T$ and $\Delta$ approach zero} \label{entropy}
\end{figure}

\narrowtext\emph{Anti-thermalization by quantum interference.---}We have
shown that when $\gamma_3\neq 0$, in the dynamic thermalization of the
three-level system all initial information will finally be erased. As
mentioned above, things are not so simple when $\gamma_3=0$. In fact, the
steady state of the $\Lambda$-type three-level system immersed in a zero
temperature bath will depend on its initial state if the bath coupling
between the two lower levels is forbidden. This phenomenon is referred to as
anti-thermalization.

When $\gamma_3=0$, the analysis of the $\Lambda$-type three level system
immersed in a heat bath can be made in the same way as presented above. The
time evolution is described by the master equation~(\ref{mastereqaution})
with $\gamma _{3}=0$. At zero temperature, we have $\det (\mathbf{M})=0$.
Thus the optical Bloch equation is reduced to
\begin{eqnarray}  \label{Blocheqforlambdaatom}
\langle \dot{\hat{\sigma}}_{ee}\rangle &=&-(\gamma _{1}+\gamma _{2})\langle
\hat{\sigma}_{ee}\rangle ,\hspace{0.5cm}\langle \dot{\hat{\sigma}}
_{g_{l}g_{l}}\rangle =\gamma _{l}\langle \hat{\sigma}_{ee}\rangle,  \notag \\
\mathtt{Re}[\langle \dot{\hat{\sigma}}_{g_{2}g_{1}}\rangle] & =&\frac{1}{2}%
(\gamma _{12}+\gamma _{21})\langle \hat{\sigma}_{ee}\rangle -\Delta \mathtt{%
Im}[\langle \hat{\sigma}_{g_{2}g_{1}}\rangle], \\
\mathtt{Im}\langle \dot{\hat{\sigma}}_{g_{2}g_{1}}\rangle & =&\Delta \mathtt{%
Re}[\langle \hat{\sigma}_{g_{2}g_{1}}\rangle],\hspace{0.1 cm}\langle \dot{%
\hat{\sigma}}_{eg_{l}}\rangle =-\frac{1}{2}(\gamma _{1}+\gamma _{2})\langle
\hat{\sigma}_{eg_{l}}\rangle,  \notag
\end{eqnarray}
where $l=1,2$.

These equations can be solved straightforwardly to obtain the transient
solutions for $\langle \hat{\sigma}_{ee}(t)\rangle ,\langle \hat{\sigma}%
_{g_{l}g_{l}}(t)\rangle $ and $\langle \hat{\sigma}_{g_{2}g_{1}}(t)\rangle $
where $l=1,2$. From these transient solutions it follows that
\begin{eqnarray}
\langle \hat{\sigma}_{g_{l}g_{l}}\rangle _{ss}&=&\langle \hat{\sigma}%
_{g_{l}g_{l}}(0)\rangle +\frac{\gamma _{l}}{\gamma _{1}+\gamma _{2}}\langle
\hat{\sigma}_{ee}(0)\rangle, \\
\mathtt{Re}[\langle \hat{\sigma}_{g_{2}g_{1}}\rangle _{ss}]& =&\mathtt{Re}%
[\langle \hat{\sigma}_{g_{2}g_{1}}(0)\rangle] +\frac{(\gamma _{12}+\gamma
_{21})}{2(\gamma _{1}+\gamma _{2})}\langle \hat{\sigma}_{ee}(0)\rangle,
\notag  \label{steadystatesolutionlambda}
\end{eqnarray}
and $\langle \hat{\sigma}_{ee}\rangle _{ss}=\langle \hat{\sigma}
_{eg_{1}}\rangle _{ss}=\langle \hat{\sigma}_{eg_{2}}\rangle _{ss}=0$ , $%
\mathtt{Im}[\langle \hat{\sigma}_{g_{2}g_{1}}\rangle _{ss}]=\mathtt{Im}%
[\langle \hat{\sigma}_{g_{2}g_{1}}(0)\rangle]$, where $l=1,2$. This is just
the steady state solution of equation~(\ref{Blocheqforlambdaatom}).

Equations~(\ref{steadystatesolutionlambda}) clearly show that the steady
state of the $\Lambda $-type three-level system depends on its initial
state. In the steady state, the decaying probabilities from the excited
state $|e\rangle$ to the ground states $|g_{1}\rangle$ and $|g_{2}\rangle $
are respectively $\gamma _{1}/(\gamma _{1}+\gamma _{2})$ and $\gamma
_{2}/(\gamma _{1}+\gamma _{2})$. Moreover, as is shown in Eq.~(\ref%
{steadystatesolutionlambda}), in the present case the dynamic thermalization
will preserve or even increase the off-diagonal elements of the density
matrix of the initial state while in the previous case with $\gamma_3\neq 0$%
, the dynamic thermalization will lead the system to a steady state whose
density matrix possesses no off-diagonal elements. It is also noticed that
the steady state of the three-level system is independent of the initial
off-diagonal elements between $|e\rangle $ and $|g_{l}\rangle $ ($l=1,2$),
and when it is initially prepared in the superposition state of the two
ground states its final steady state will be the same as the initial state.
Finally let us present two simple examples. Take $\gamma
_{1}=\gamma_{2}=\gamma _{12}=\gamma _{21}=\gamma $ and $\Delta =0$, then for
the initial state $|\psi (0)\rangle =|e\rangle $ we have the steady state $%
|\psi (\infty )\rangle =(|g_{1}\rangle +|g_{2}\rangle )/\sqrt{2}$ and for
the initial state $|\psi (0)\rangle =|g_{1}\rangle $ we have the steady
state $|\psi (\infty )\rangle =|g_{1}\rangle $.

\narrowtext\emph{Summary.---}In summary, in this letter the SSB effect in
dynamic thermalization is studied through a three-level system immersed in a
heat bath inducing cycle transition couplings. Careful calculation is
carried out from the master equation approach to examine the thermalization
dynamics when the temperature approaches zero. By this investigation it is
concluded that when there is no selection rule to forbid any one of the bath
induced cycle transition couplings, the canonical thermal state can be
reached as a steady state solution of the master equation at zero
temperature and if the bath induced transition between the two lower
(higher) energy states of $\Lambda $-type ($V$-type) atom is forbidden the
anti-thermalization phenomenon will happen due to the quantum interference
between the transition from the lower state and the transition to the higher
state. In this latter case, the final steady state of the three-level system
will depend on its initial state, and thus will preserve some of the initial
information. This means that the initial information of the system cannot be
completely erased by thermalization and the third law of thermodynamics does
not work in the conventional fashion.

The work is supported by National Natural Science Foundation of China  and
the National Fundamental Research Programs of China under Grant.

\end{document}